\documentclass[pra,twocolumn]{revtex4}
\usepackage{amsfonts}
\usepackage{amsmath}
\usepackage{amssymb}
\usepackage{graphicx}

\begin{document}

\title{Polarization squeezing and multipartite entanglement of triphoton
states}

\author{G. R. Jin}
\email{grjin@bjtu.edu.cn}
\author{S. Luo}
\author{Y. C. Liu}
\affiliation{Department of Physics, Beijing Jiaotong University,
Beijing 100044, China}
\author{H. Jing}
\affiliation{Department of Physics, Henan Normal University,
Xinxiang, Henan 453007, China}
\author{W. M. Liu}
\affiliation{Beijing National Laboratory for Condensed Matter
Physics, Institute of Physics, Chinese Academy of Sciences, Beijing
100080, China}
\date{\today }

\begin{abstract}Based upon standard angular momentum theory, we develop a
framework to investigate polarization squeezing and multipartite
entanglement of a quantum light field. Both mean polarization and
variances of the Stokes parameters are obtained analytically, with
which we study recent observation of triphoton states [L. K. Shalm,
{\it et al}, Nature \textbf{457}, 67 (2009)]. Our results show that
the appearance of maximally entangled NOON states accompanies with a
flip of mean polarization and can be well understood in terms of
quantum Fisher information.

\vskip 0.1cm


\end{abstract}

\maketitle

\section{Introduction}

Polarization squeezing and quantum entanglement of a light field
have received much attention for decades not only because of
fundamental physical interests, but also for potential applications
in quantum metrology and quantum information
\cite{Ou,Lee,Lam,Mitchell}. Formally, the squeezing is defined as a
reduction of polarization uncertainty below shot-noise limit (SNL),
which is standard quantum limit imposed by Heisenberg uncertain
relationship. It has been shown that the squeezing, closely related
to multipartite entanglement is aroused from quantum correlation
effect among individual particles
\cite{Kitagawa,Wineland1,Wineland2,Sorensen,Jin09}.

Quantum metrology based upon maximally entangled NOON states results
in super-resolving phase estimations \cite{Yurke,Giovannetti}.
However, a deterministic optical source of the entangled states is
yet to be realized due to technical difficulties \cite{Resch}. Using
various state-projection measurements, so far there are a lot of
groups have realized few-photon NOON states
\cite{Mitchell,Walther,Nagata,Guo,Lu,Shalm}. In particular, Shalm
\textit{et al.} have succeed in preparing maximally entangled NOON
state of the triphotons \cite{Shalm}. Counter-intuitively, they
found that the NOON state does not show polarization squeezing, just
like quantum uncorrelated coherent states. To explain it, in this
paper we theoretically study polarization squeezing and multipartite
entanglement of the triphotons. Analytical expressions of the
reduced and increased variances are presented to determine the
squeezing parameters and quantum Fisher information (QFI)
\cite{Caves,Wootters,Smerzi,WangX2} of the triphoton states. Our
results show that the mean polarization of the triphotons changes
its sign with the appearance of NOON states.

The outline of this paper is arranged as follows: in Sec. 2, we give
notations and definitions of the polarization squeezing followed by
a rigorous analytical approach for obtaining the reduced and the
increased variances of Stokes parameters. Sec. 3 is devoted to
consider the maximally entangled triphoton state. Firstly, we solve
the variances and the optimal squeezing direction for the
polarization squeezed state. To proceed, we investigate the
relationship between the squeezing and the entanglement. Our
conclusion will be presented in Sec. 4.

\section{Theoretical background}

In analogy with classical optics, the polarization a light field can
be described by Stokes vectors $(c=\hbar =1)$ \cite{Duan}
\begin{eqnarray}
&&\hat{S}_{0}=\frac{1}{2}(\hat{a}_{H}^{\dagger }\hat{a}_{H}+\hat{a}
_{V}^{\dagger }\hat{a}_{V}),
\hat{S}_{1}=\frac{1}{2}(\hat{a}_{H}^{\dagger}
\hat{a}_{H}-\hat{a}_{V}^{\dagger }\hat{a}_{V}),  \nonumber \\
&&\hat{S}_{2}=\frac{1}{2}(\hat{a}_{H}^{\dagger }\hat{a}_{V}+\hat{a}
_{V}^{\dagger }\hat{a}_{H}),
\hat{S}_{3}=\frac{1}{2i}(\hat{a}_{H}^{\dagger}
\hat{a}_{V}-\hat{a}_{V}^{\dagger }\hat{a}_{H}),
\label{stocksoperator}
\end{eqnarray}%
where $\hat{a}_{H,V}$ and $\hat{a}_{H,V}^{\dagger}$ are annihilation
and creation operators for the horizontal and vertical polarization
modes, respectively. The photon operators satisfy bosonic
commutation relations $[\hat{a}_{\mu },\hat{a}_{\nu }^{\dagger
}]=\delta _{\mu \nu }$, with $\mu ,\nu \in \{H,V\}$. The stokes
vectors $\hat{S}_{1}$, $\hat{S}_{2}$, and $\hat{S}_{3}$ obey SU(2)
algebra: $[\hat{S}_{i},\hat{S}_{j}]=i\hat{S}_{k}$, with $i,j,k\in
\{1,2,3\}$, corresponding to horizontally, linearly at $ 45^{\circ
}$, and right-circularly polarized axes \cite{Lam}, respectively.
For a fixed photon number $N$ ($=2s$),
$\hat{S}^{2}=\hat{S}_{1}^{2}+\hat{S}_{2}^{2}+\hat{S}_{3}^{2}=s(s+1)$
and $\hat{S}_{0}=s$ are invariant and commutes with other three
Stokes operators. Following standard theory of angular momentum, we
choose eigenstates of $\hat{S}_{1}$, $|s,n\rangle =|s+n,s-n\rangle
_{H,V}$ as the basis of total Hilbert space, where the photon number
states are defined as usual, $|m,n\rangle _{H,V}=(a_{H}^{\dagger
})^{m}(a_{V}^{\dagger })^{n}|0\rangle /\sqrt{m!n!}$. The SU(2)
angular momentum states obey $\hat{S}_{\pm }|s,n\rangle =\sqrt{
(s\mp n)(s\pm n+1)}|s,n\pm 1\rangle $, with the ladder operators
$\hat{S}_{\pm }=\hat{S}_{2}\pm i\hat{S}_{3}$.

Any quantum polarization state $|\Psi \rangle$ is characterized by
the mean polarization $\langle \vec{S}\rangle =(\langle
\hat{S}_{1}\rangle ,\langle \hat{S}_{2}\rangle ,\langle
\hat{S}_{3}\rangle )$ on a Poincar\'{e} sphere, where $\langle
\hat{S}_{i}\rangle =\langle \Psi |\hat{S}_{i}|\Psi \rangle $ for
$i=1$, $2$, $3$. To define the concept of polarization squeezing, it
is convenient to introduce three orthogonal polarization vectors
$\hat{S}_{n_{i}}=\hat{S}\cdot \hat{n}_{i}$, with
\begin{eqnarray}
\hat{n}_{1} &=&(0,-\sin \phi ,\cos \phi ),  \nonumber \\
\hat{n}_{2} &=&(\sin \theta ,-\cos \theta \cos \phi ,-\cos \theta
\sin \phi
),  \label{n1n2n3} \\
\hat{n}_{3} &=&(\cos \theta ,\sin \theta \cos \phi ,\sin \theta \sin
\phi ), \nonumber
\end{eqnarray}%
where the polar angle $\theta$ and the azimuth angle $\phi$ obey
$\sin \theta =r/|\langle \vec{S}\rangle |$, $\cos \theta =\langle
\hat{S}_{1}\rangle /|\langle \vec{S}\rangle |$, $\sin \phi =\langle
\hat{S}_{3}\rangle /r$, $\cos \phi =\langle \hat{S}_{2}\rangle /r$.
Here, $|\langle \vec{S}\rangle |=(\langle \hat{S}_{1}\rangle
^{2}+\langle \hat{S}_{2}\rangle ^{2}+\langle \hat{S}_{3}\rangle
^{2})^{1/2}$, denoting the length of the mean polarization, and
$r=(\langle \hat{S}_{2}\rangle ^{2}+\langle \hat{S}_{3}\rangle
^{2})^{1/2}=|\langle \vec{S}\rangle |\sin \theta $. Note that the
mean polarization $\langle \vec{S}\rangle =|\langle \vec{S}\rangle
|\hat{n}_{3}$ with its length $|\langle \vec{S}\rangle |=\langle
\hat{S}_{n_{3}}\rangle $ \cite{Jin09}. The two orthogonal
polarization vectors normal to $\langle \vec{S}\rangle $ (i.e.,
$\hat{n}_{3}$) satisfy Heisenberg uncertainty relationship
\begin{equation}
(\Delta \hat{S}_{n_{1}})^{2}(\Delta \hat{S}_{n_{2}})^{2}\geq
\frac{1}{4}|\langle \hat{S}_{n_{3}}\rangle |^{2},
\end{equation}%
where $(\Delta \hat{S}_{n_{i}})^{2}\equiv \langle
\hat{S}_{n_{i}}^{2}\rangle -\langle \hat{S}_{n_{i}}\rangle ^{2}$,
denoting the variance of the Stokes operator $\hat{S}_{n_{i}}$ for
$i=1$, $2$. It is well known that the minimal uncertainty
relationship is obtained for SU(2) coherent state
\begin{equation}
\left\vert \theta ,\phi \right\rangle =e^{-i\theta \hat{S}%
_{n_{1}}}\left\vert s,s\right\rangle =e^{i\theta (\hat{S}_{2}\sin
\phi -\hat{S}_{3}\cos \phi )}\left\vert s,s\right\rangle ,
\end{equation}%
which is also an eigenstate of $\hat{S}_{n_{3}}$ with eigenvalue $s$
(i.e., $\langle \hat{S}_{n_{3}}\rangle =|\langle \vec{S}\rangle
|=s$), and thereby $(\Delta \hat{S}_{n_{1}})^{2}=(\Delta
\hat{S}_{n_{2}})^{2}=s/2$. Here, the value $s/2$ is termed as
standard quantum limit, or the shot-noise limit (SNL). The squeezing
is defined if any polarization component normal to $\langle
\vec{S}\rangle$ has a reduced variance below the SNL
\cite{Kitagawa}. It is obvious to choose the squeezed polarization
component as
\begin{equation}
\hat{S}_{\gamma }=\hat{S}\cdot \hat{n}_{\gamma }=\hat{S}_{n_{1}}\cos
\gamma+\hat{S}_{n_{2}}\sin \gamma ,  \label{S_gamma}
\end{equation}%
where $\gamma $ is arbitrary angle with respect to $\hat{n}_{1}$.
Due to the relation $\langle \hat{S}_{\gamma }\rangle =0$, the
variance of $\hat{S}_{\gamma}$ takes the form $(\Delta
\hat{S}_{\gamma })^{2}=[C+A\cos (2\gamma )+B\sin (2\gamma )]/2$,
where $A=\langle \hat{S}_{n_{1}}^{2}-\hat{S} _{n_{2}}^{2}\rangle$,
$B=\langle \hat{S}_{n_{1}}\hat{S}_{n_{2}}+\hat{S}
_{n_{2}}\hat{S}_{n_{1}}\rangle$, and $C=\langle
\hat{S}_{n_{1}}^{2}+\hat{S}_{n_{2}}^{2}\rangle$. Minimizing $(\Delta
\hat{S}_{\gamma })^{2}$ with respect to $\gamma$, we get
\cite{Jin09,Jin07,Jin07PRA}
\begin{equation}
V_{\pm }=\frac{1}{2}\left[C\pm \sqrt{A^{2}+B^{2}}\right] ,
\label{Vpm}
\end{equation}%
where the reduced variance $V_{-}=\min_{\gamma}(\Delta
\hat{S}_{\gamma })^{2}$, denoting the optimal squeezing along
$\hat{n}_{\gamma }$ with $\gamma =\gamma_{\text{op}}\equiv \lbrack
\pi +\arctan (B/A)]/2$; while the increased variance
$V_{+}=\max_{\gamma }(\Delta \hat{S}_{\gamma })^{2}$, corresponding
to the anti-squeezing along $\hat{n}_{\gamma }$ with $\gamma =\pi
/2+\gamma_{\text{op}}$. Remarkably, the above analysis provide us
explicit form of the optimal squeezing angle $\gamma_{\text{op}}$
and that of $V_{\pm }$. For the minimal uncertainty state $|\theta
,\phi \rangle $, $V_{+}=V_{-}=s/2$, so the inequality
\begin{equation}
\xi ^{2}=\frac{2(V_{-})}{s}<1
\end{equation}
recognizes polarization squeezed states \cite{Lius}. In the
following, we will study polarization squeezing and entanglement of
the triphotons, which has been demonstrated recently by Shalm et al.
\cite{Shalm}.

\section{The triphoton states}

Recently, Shalm et al. \cite{Shalm} have succeed in preparing the
triphoton states. Due to the lack of ideal single-photon sources,
they adopted type-II spontaneous parametric downconversion (SPDC)
and an attenuated laser (a local oscillator, LO). A pair of
orthogonally polarized photons from the SPDC and a single photon
from the LO are overlapped and placed into the same mode to produce
a state likes $\hat{a}_{45}^{\dag }\hat{a}_{-45}^{\dag }a_{H}^{\dag
}|0\rangle =(\hat{a}_{H}^{\dagger 2}-\hat{a}_{V}^{\dagger
2})\hat{a}_{H}^{\dag }|0\rangle $. This state is then sent to a
variable partial polarizer (VPP),
with which one can manipulate the polarization of light to produce \cite%
{Shalm}:
\begin{equation}
|\Psi \rangle _{T}\propto e^{-\hat{S}_{1}\ln
(T)}(\hat{a}_{H}^{\dagger 2}-\hat{a}_{V}^{\dagger
2})\hat{a}_{H}^{\dagger }|0\rangle ,
\end{equation}%
where $T=T_{V}/T_{H}$, denoting the transmissivity ratio of the
horizontal and the vertical modes photons. In the basis of
$\{|s,n\rangle \}$ with $s=3/2$, the polarization state can be
rewritten as $|\Psi \rangle
_{T}=(3+T^{4})^{-1/2}[3^{1/2}|3/2,3/2\rangle -T^{2}|3/2,-1/2\rangle
]$. If the VPP is tuned to transmit only the horizontal polarized
photons (i.e., $T=0)$, it becomes $|\Psi
\rangle_{T}=|3/2,3/2\rangle=|3, 0\rangle_{H,V}$, corresponding to a
coherent state $|\theta, \phi \rangle $ with $\theta =0$. Utilizing
the VPP, it is now possible to tune the ratio $T$ from $0$ up to
$1.8$ in the experiment \cite{Shalm}.

\begin{figure}[htbp]
\centerline{\includegraphics[width=8cm,
angle=0]{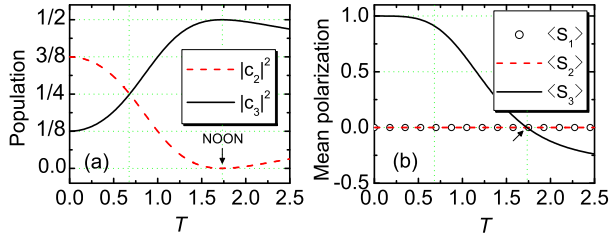}}\caption{(Color online) (a) Population
distribution $|c_{2}|^{2}$ (red dash) and $|c_{3}|^{2}$ (solid) as a
function of the transmissivity ratio $T$. (b) Normalized mean
polarization $\frac{\langle \vec{S}\rangle}{s} =\frac{1}{s}(\langle
\hat{S}_{1}\rangle, \langle \hat{S}_{2}\rangle, \langle
\hat{S}_{3}\rangle)$. Vertical grid lines in (a) and (b) denote
$T\simeq 0.7$ and $T=\sqrt{3}\simeq 1.7$, respectively.}
\label{fig1}
\end{figure}

After the VPP, a quarter-wave plate (QWP) is adopted to rotate the
polarization state into the basis of $\hat{S}_{3}$ \cite{Shalm}. The
action of the QWP can be described formally by an unitary
transformation $\exp (i\frac{\pi }{2}\hat{S}_{2})$, which in turn
leads to a kind of triphoton states
\begin{eqnarray}
|\Psi \rangle&=&c_{2}(i|2,1\rangle _{H,V}-|1,2\rangle _{H,V})  \nonumber \\
&&+c_{3}(|3,0\rangle _{H,V}-i|0,3\rangle _{H,V}),  \label{triphoton}
\end{eqnarray}%
where the $T$-dependent probability amplitudes
\begin{eqnarray}
c_{2}&=&\frac{1}{2\sqrt{2}} \frac{3-T^2}{\sqrt{3+T^{4}}},
c_{3}=\frac{1}{2}\sqrt{\frac{3}{2}}\frac{1+T^2}{\sqrt{3+T^4}}.
\label{amplitudes}
\end{eqnarray}
In Fig.~\ref{fig1}(a), we plot population distributions
$|c_{2}|^{2}$ and $|c_{3}|^{2}$ as a function of $T$ . It is found
that (i) $c_{3}=c_{2}/\sqrt{3}=1/(2\sqrt{2})$ at $T=0$; (ii) $
c_{3}=c_{2}=1/2$ at $T=3^{1/4}(2-\sqrt{3})^{1/2}\simeq 0.7$; (iii)
$c_{2}=0$ and $c_{3}=1/\sqrt{2}$ at $T=\sqrt{3}\simeq 1.7$. The
first case represents a coherent state $|\theta =\pi /2,\phi =\pi
/2\rangle$, obtained from $|3,0\rangle_{H,V}$ through a rotation of
$\pi /2$ angle about $-\hat{S}_{2}$ axis, just a result of the QWP.
The second case corresponds to a phase state with equal populations,
and the latter case represents the \textquotedblleft NOON" (i.e.,
Greenberger-Horne-Zeilinger, or Schr\"{o}dinger ``cat") state:
$|\Psi \rangle =\frac{1}{\sqrt{2}}(|3,0\rangle _{H,V}-i|0,3\rangle
_{H,V})$. Such a kind of maximally entangled triphoton state has
been demonstrated by Shalm et al. \cite{Shalm}.

\subsection{Polarization squeezing of the triphotons}

Based upon Eq. (\ref{triphoton}), we begin to investigate the
polarization squeezing of the triphotons. Firstly, we have to
determine the mean polarization $\langle \vec{S}\rangle =(\langle
\hat{S}_{1}\rangle ,\langle \hat{S}_{2}\rangle ,\langle
\hat{S}_{3}\rangle )$, where $\langle \hat{S}_{2}\rangle
=$Re$\langle \hat{S}_{+}\rangle $ and $\langle \hat{S} _{3}\rangle
=$Im$\langle \hat{S}_{+}\rangle $. Because of \emph{imaginary} value
of $\langle \hat{S}_{+}\rangle $, we have $\langle
\hat{S}_{1}\rangle =\langle \hat{S}_{2}\rangle =0$ [see
Fig.~\ref{fig1}(b)], i.e., the mean polarization being parallel with
the $\hat{S}_{3}$ axis. In this case, Eq.(\ref{S_gamma}) reduces to
$\hat{S}_{\gamma}=-\hat{S}_{2}\cos \gamma +\hat{S} _{1}\sin \gamma
$, where we have set $\theta =\phi =\pi /2$ in Eq.(\ref{n1n2n3}).
The coefficients given in Eq. (\ref{Vpm}) can be solved analytically
as
\begin{eqnarray}
A &=&\langle \hat{S}_{2}^{2}-\hat{S}_{1}^{2}\rangle
=\frac{15}{8}-\frac{
3(9c_{3}^{2}+c_{2}^{2})+8\sqrt{3}c_{3}c_{2}}{4}, \nonumber \\
C &=&\langle \hat{S}_{2}^{2}+\hat{S}_{1}^{2}\rangle
=\frac{15}{8}+\frac{9c_{3}^{2}+c_{2}^{2}-8\sqrt{3}c_{3}c_{2}}{4},
\label{AC}
\end{eqnarray}
where probability amplitudes $c_2$ and $c_3$ are given by
Eq.(\ref{amplitudes}). Since $A<0$ and $B=-\langle
\hat{S}_{2}\hat{S}_{1}+\hat{S}_{1}\hat{S}_{2}\rangle =0$, we get the
optimal squeezing angle $\gamma_{\text{op}}=\frac{1}{2}[\pi +\arctan
(B/A)]=\pi $. Without any ambiguous, our analytic results show that
the optimal polarization squeezing is along $\hat{S}_{2}$ and the
anti-squeezing along the $\hat{S}_{1}$ axis.

\begin{figure*}
\centerline{\includegraphics[width=10cm,angle=0]{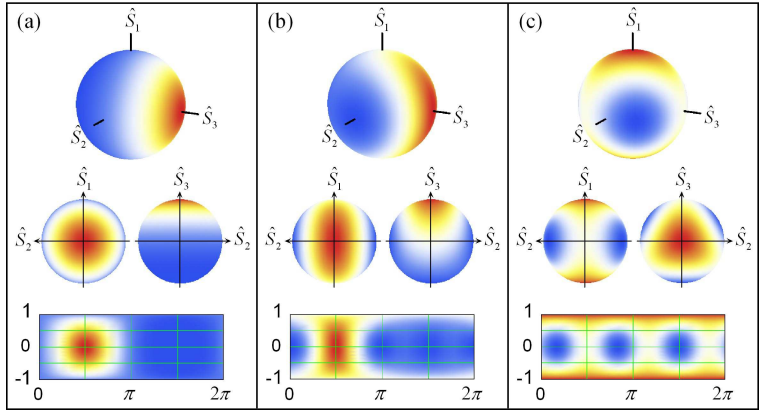}}
\caption{(Color online) Quasi-probability distribution (Husimi Q
function) $Q(\theta, \phi)$ for SU(2) coherent state at $T=0$ (a),
polarization squeezed state at $T=1$ (b), and maximally entangled
NOON state at $T=\sqrt{3}\simeq 1.7$ (c). The top two plots are the
Q function on the Poincar\'{e} sphere, and the bottom plots are the
Q function in a two-dimensional phase space ($\phi$, $p$) with
$0\leq\phi<2\pi$ and $-1\leq p\leq1$. The vertical coordinate
$p=\cos\theta$, denoting population imbalance between the
horizontally and the vertically polarized photons. Red (blue)
shading represents a larger (smaller) value of the Q function.}
\label{fig2}
\end{figure*}

Evolution of polarization uncertainty can be illuminated
schematically in terms of quasi-probability distribution. As shown
in Fig.~\ref{fig2}, we plot the Husimi Q function on Poincar\'{e}
sphere, which is defined as the expectation value of the density
matrix operator $\hat{\rho}$ with respect to the SU(2) coherent
states \cite{Kitagawa}:
\begin{equation}
Q(\theta ,\phi )=\langle \theta ,\phi |\hat{\rho}|\theta ,\phi
\rangle . \label{Qfuncdefine}
\end{equation}%
In comparison with Wigner function \cite{Agarwal,Dowling94}, the Q
function is regular, positive definite, and especially suitable for
illuminations \cite{Trimborn}. From Fig.~\ref{fig2}(a), we find that
the triphoton state at $T=0$ shows an isotropic quasi-probability
distribution because of the minimum uncertainties of two orthogonal
polarization components normal to $\langle \vec{S}\rangle$. When
$T=1$, the density of $Q(\theta ,\phi)$ becomes an elliptic shape
[Fig.~\ref{fig2}(b)] due to the squeezing and the anti-squeezing
along $\hat{S}_{2}$ and $\hat{S}_{1}$, respectively. Via a mapping
$p=\cos\theta$, the Husimi Q function can be also plotted in a
two-dimensional phase space ($\phi, p$) \cite{Trimborn,Jin08}. For
$T=0$ and $1$ cases, maximal values of $Q(\phi, p)$ appear at
$\phi=\pi/2$ and $p=\cos\theta=0$. Inserting $\theta =\phi=\pi/2$
into the last expression of Eq.(\ref{n1n2n3}), one can find that the
direction of the mean polarization is in fact along $\hat{S}_{3}$.
It is shown from Fig.~\ref{fig2}(c) that the NOON state at
$T=\sqrt{3}\simeq 1.7$ shows a threefold symmetrical
quasi-probability distribution \cite{Shalm}, with its density peaked
at the north and the south poles of the $\hat{S}_{1}$ axis [$p=\cos
\theta =\pm1$], which indicates all the photons being either
horizontally polarized or vertically polarized. To somewhat
counter-intuitively, however, such a maximally entangled state shows
the reduced variance $V_{-}$ equal to the SNL [i.e., $2V_{-}/s=0$dB,
see Fig.~\ref{fig3}(a)], just like quantum uncorrelated coherent
states. How to identify the NOON state becomes a subtle but
important problem.

\subsection{Multipartite entanglement of the triphotons}

Shalm et al \cite{Shalm} provide a transparent experimental results
to test the relationship between spin squeezing and multipartite
entanglement. Previously, it has been proposed that a useful
squeezing for quantum metrology and quantum entanglement obeys:
$\zeta_{\text{S}}^{2}=2s(V_{-})/|\langle \vec{S}\rangle
|^{2}=(s/|\langle \vec{S}\rangle|)^{2}\xi^{2}<1$ \cite{Sorensen}.
Rather than it, Pezz\'{e} and Smerzi \cite{Smerzi} recently proposed
a more general criterion for $N$-particle entanglement:
\begin{equation}
\chi^{2}=\frac{N}{F[\hat{\rho},\hat{S}_{\gamma }]}<1,  \label{chi2}
\end{equation}%
where $F[\hat{\rho},\hat{S}_{\gamma }]=4V_{+}$, denoting quantum
Fisher information for a pure state $\hat{\rho}=|\Psi \rangle
\langle \Psi |$ \cite {Caves}, and thereby $\chi ^{2}=s/(2V_{+})$.
According to Ref. \cite{Smerzi}, Eq. (\ref{chi2}) is not only a
sufficient condition for multipartite entanglement, but also a
sufficient and necessary condition for sub-shot-noise phase
estimation. Moreover, it has been shown that the criterion
$\chi^2<1$ can be used to distinguish and characterize quantum
critical behaviors of the Lipkin-Meskhov-Glick model \cite{WangX2}.

\begin{figure}[htbp]
\centerline{\includegraphics[width=8cm,angle=0]{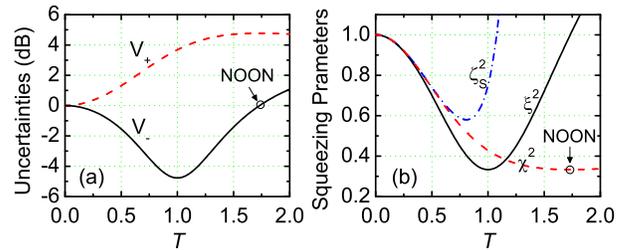}}\caption{(Color
online) (a) The reduced (solid) and the increased (red dash)
uncertainties $V_{\pm}$ relative to the SNL, $s/2$. (b) The
squeezing parameters $\zeta_{\text{S}}^{2}$ (blue dot-dash),
$\xi^{2}$ (solid), and $\chi^{2}$ (red dash) as a function of the
transmissivity ratio $T$. The maximal squeezing with
$\xi_{\min}^2=2(V_{-})_{\min}/s=1/3$ (i.e., $-4.77$dB) is obtained
at $T=1$; however, the maximal entangled NOON state appears at
$T=\sqrt{3}>1$ [see also Fig.~\ref{fig1}(a)], which shows
$\chi^2_{\min}=s/[2(V_{+})_{\max}]=1/3$.} \label{fig3}
\end{figure}

Substituting Eq. (\ref{AC}) into Eq. (\ref{Vpm}), we obtain the
anti-squeezed variance $V_{+}=(\Delta
\hat{S}_1)^2=\frac{1}{2}(9c_{3}^{2}+c_{2}^{2})$, and the squeezed
variance:
\begin{equation}
V_{-}=(\Delta
\hat{S}_2)^2=\frac{1}{4}\left[\frac{15}{2}-(9c_{3}^{2}+c_{2}^{2}+8\sqrt{3}c_{3}c_{2})\right],
\label{VpVm}
\end{equation}
where the probability amplitudes $c_2$ and $c_3$ are given by Eq.
(\ref{amplitudes}). In Fig.~\ref{fig3}, we plot $V_{\pm}$,
$\zeta_{\text{S}}^{2}$, $\xi^{2}$, and $\chi^{2}$ as a function of
the transmissivity ratio $T$. For the SU(2) coherent state at $T=0$,
the mean spin $|\langle \vec{S}\rangle|=s$ [see Fig.~\ref{fig1}(b)]
and the variances $V_{+}=V_{-}=s/2$ (i.e., $0$ dB), which yields
$\zeta_{\text{S}}^{2}=\xi ^{2}=\chi ^{2}=1$. With the increase of
$T$ up to $1$, the variance $V_{-}$ decreases along $\hat{S}_{2}$ at
the expense of an increased variance $V_{+}$ along $\hat{S}_{1}$. As
shown in Fig.~\ref{fig3}(a) and (b), the squeezed variance $V_{-}$
and also $\xi^{2}$ monotonically decrease to its minimum value $\xi
_{\min}^{2}=2(V_{-})_{\min }/s=1/3$ ($\sim -4.77$dB), corresponding
to a maximally squeezed state at $T=1$. However, the squeezing
parameter $\zeta_{\text{S}}^{2}$ reaches to its minimal value ($\sim
0.58$) at $T\sim 0.81$ [see blue dot-dash curve of Fig.~\ref{fig3}
(b)]. This is because of the reduced mean polarization $\langle
\vec{S}\rangle <s$ and different evolution rates between $V_{-}$ and
$\langle \vec{S}\rangle$ \cite{Jin09}. After $T=1$, both $V_{+}$ and
$V_{-}$ begin to increase due to the so-called over-squeezing
\cite{Shalm}. As mentioned above, the NOON state appears at
$T=\sqrt{3}$, where the squeezing parameters $\xi^{2}=1$ and
$\zeta_{\text{S}}^{2}\rightarrow \infty$ due to $|\langle
\vec{S}\rangle |\rightarrow 0$. From red dash line of
Fig.~\ref{fig3}(b), one can find that $\chi^{2}$ continuously
deceases from $1$ for the coherent state to the smallest value $1/3$
for the NOON state. The appearance of the NOON state accompanies
with a flip of the mean polarization [see Fig.~\ref{fig1} (b)]. Such
a result keeps hold for $N$-photon NOON state:
\begin{equation}
|\Psi\rangle_{\text{NOON}}=\frac{1}{\sqrt{2}}(|N,0\rangle
_{H,V}+e^{i\varphi}|0,N\rangle _{H,V}),
\end{equation}
where $\varphi$ is an arbitrary phase. Note that the NOON state
exhibits vanishing mean polarization $\langle \vec{S}\rangle=0$ and
the largest photon number fluctuation between horizontal and
vertical modes $V_{+}=(\Delta \hat{S}_1)^2=s^2$ \cite{Jin10}.
Moreover, for $|\Psi\rangle_{\text{NOON}}$, the reduced variance
$V_{-}=(\Delta \hat{S}_2)^2=s/2$, which is finite and equal to the
SNL, so $\zeta_{\text{S}}^{2}\rightarrow \infty$, $\xi^{2}=1$, and
$\chi^{2}=1/N$.

\section{Conclusion}

In summary, we investigated theoretically polarization squeezing and
multipartite entanglement of the triphoton states. Analytical
expressions of the reduced and increased variances $V_{\pm}$ of the
Stokes parameters are presented by using standard angular momentum
theory. As two {\it different} nonclassical effects, we find that
polarization squeezing and bipartite entanglement of the triphotons
can be measured respectively, by the parameters $\xi^{2}=2(V_{-})/s$
and $\chi^{2}=s/[2(V_{+})]$. In particular, recent experimental
observations of the NOON state \cite{Shalm} can be well understood
in terms of the entanglement parameter $\chi^2$, which deceases
monotonically from the shot-noise limit $1$ for the coherent state,
to the so-called Heisenberg limit $1/N$ (here $N=3$) for the
maximally entangled NOON state.

\section*{Acknowledgments}

We thank Profs. X. Wang, C.~P. Sun, S. Yi, and L. You for helpful
discussions. This work is supported by the NSFC (Contract
No.~10804007) and the SRFDP (Contract No.~200800041003). S.L. and
Y.C.L. are supported by National Innovation Experiment Program for
University Students (BJTU No.~091000438). W.M.L. is supported by the
NSFC (No.~60525417 and No.~10874235), and the NKBRSFC
(No.~2006CB921400 and No.~2009CB930700).


\end{document}